\documentclass{article}
\usepackage[pdftex]{graphicx}
\usepackage{epstopdf}
\usepackage{latexsym}
\usepackage{amssymb}
\begin{document}
\title{Flow of autonomous traffic on a single multi-lane street}
\author{Federico Polito and Fergal Dalton\\
Consiglio Nazionale delle Ricerche, Istituto dei
Sistemi Complessi,\\ sede di Tor Vergata, Via del Fosso del
Cavaliere 100, 00133 Roma, Italy.}

\date{\today}

\maketitle

\begin{abstract}

We investigate the behaviour of an original traffic model.
The model considers a single multi-lane street, populated
by autonomous vehicles directed from either end to the other.
Lanes have no intrinsic directionality, and the vehicles are
inserted at random at either end and any lane.  Collision
avoidance is fully automatic and, to enhance the transport
capacity of the street, vehicles form {\it trains} in which
they may travel at high speed quite close to the vehicle
in front.

We report on the transit times for vehicles under a wide
variety of conditions:
vehicle insertion probability \& imbalance and their
maximum speed distribution.  We also outline an
interesting feature of the model, that the complex
interactions of many vehicles are considerably more
powerful than a simple "keep left" directive which
each vehicle should obey.


\end{abstract}


\section*{Motivations}

\begin{figure}
\centering 
\includegraphics[scale=0.9]{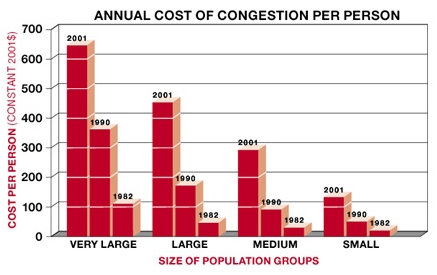}
\caption{\label{transportcost} The increase in the real cost of transport from 1982 to 2001.
The figure only considers cost due to {\it wasted time} and not overall transit time from
source to destination.  Source: reference~\cite{urbanmobility}.  }
\end{figure}

\begin{figure}
\centering 
\includegraphics[scale=1]{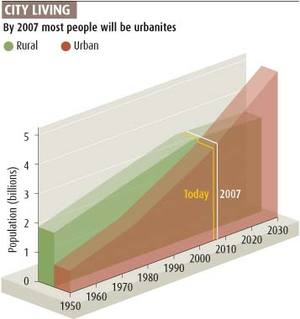}
\caption{\label{urbanliving} The number of people living in urban and rural settings worldwide.
This year (2007) urban dwellers were projected to exceed rural. Source: reference~\cite{urbanliving}.  }
\end{figure}

The worldwide transport infrastructure is becoming overloaded.
In urban settings there is simply insufficient available real estate
with which to increase capacity leading to increased commuting
times and costs across all cities~\cite{urbanmobility}. Figure~\ref{transportcost}
indicates how the cost of transport has increased in the last 25 years
(in constant 2001 US\$).   It has been forecast
that by this year, half the world's population will live in urban
areas (figure~\ref{urbanliving}).  Gridlock has become such an extensive
problem that London city has introduced a heavy {\it congestion
charge} for access to the city centre, though its effects are
disputed~\cite{london-charge}.  Today's cars are consuming
resources at an ever-increasing rate and, despite the ubiquitous
warnings of climate change associated with non-renewable
energy consumption (the debate is still open, but the messages are
certainly very present)~\cite{newscientist-ipcc-report},
it seems that personal transport is not yielding.

\begin{figure}
\centering 
\includegraphics[scale=0.5]{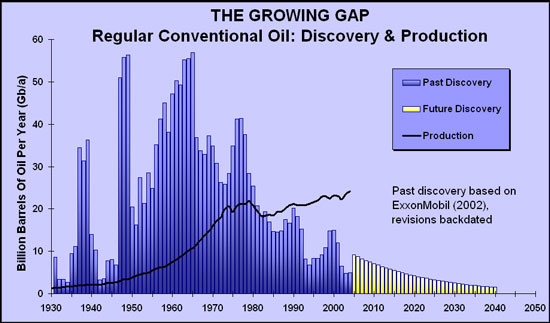}
\caption{\label{oilsupply} The history and projected future of oil reserves discovery and production.
Source: reference~\ref{oilsupply}.  }
\end{figure}

Today's transport is completely dependent on a steady, secure and
cheap supply of oil; this is difficult to guarantee for the future due
to political instability and potentially limited resources (see
figure~\ref{oilsupply}).  Indeed, some
experts have stated that the era of cheap oil is drawing to a close~\cite{hirsch-report}
(though opinion is very divided) and
that soon the world will have to turn to intrinsically difficult and
expensive oil-extraction methods (shale oil, etc).  The 
socio-economic implications are extremely profound and potentially dire.

Even renewable energy sources do not come close to solving the problem;
a drastic worldwide push started today
to develop renewable energy sources might
already be too late to substitute oil before the economic consequences
of a restricted oil supply would be felt and it is not clear that
renewable energy could ever satisfy demand.

Safety and security in transport is also a continuing problem.  The number
of deaths on our roads is increasing as cars become faster - ironically,
increased safety for the occupants implies a reduced safety for those outside.
Driver frustration with congestion is also playing a part in agressive driving.

Thus, it is becoming clear
that current trends are environmentally, economically and socially
unsustainable in the long term.  A paradigm-shift
is required to break the current trend and give a new face to transport.

\section*{Proposed Model}

In this article we report on the results of
a first 'stepping stone' to a complete computer
model of a fleet of autonomous transport units.  The overall image is of
a transport infrastructure consisting of existing roads (implanted with RFID
transponders), elevated or surface monorails and (possibly) a network of interconnected
subsurface tunnels.  Vehicles are completely self-controlled by on-board
computers~\cite{prt-pat}, other than to respond to a human initiated "summons"
and direct human input specifying the desired destination.
Nearby vehicles travelling on the infrastructure maintain
peer-to-peer~\cite{p2p} contact and share information~\cite{highway-info}
regarding traffic and viability,
thus permitting real-time unsupervised route planning and execution.

Such a futuristic transport architecture would offer many long term
benefits:
\begin{itemize}
\item Safer: Fully automated vehicles (no "human error");
             electronic response times to sudden dangers.
\item Smarter: Advanced satellite navigation and information systems;
               "intelligent" global response to perturbations.
\item Sustainable: Efficient engines "of the future"; solar panels could
                   effectively provide a city with a 500MW power-station.
\item Decongested: Real time propagation of traffic information and
                   dynamic re-routing.
\item Impartial: Suitable for all categories of end-user including
                 transportation of goods.
\item Greener: Energy source can be chosen by governments, be that
		oil, gas, nuclear or renewable; and would allow for
		centralised control over pollution.
\end{itemize}

Naturally any eventual implemetation of a system such as this would
be long and difficult, requiring many phases of integration with the
present transport infrastructure.  We do not presently consider
such problems, but only study how such a system might ultimately
behave {\it after all other conflicting forms of transport are
rendered obsolete}.  We note, however, that present day vehicles
could potentially be adapted for continued service.

As a typical city would consist of some $10^5-10^6$ vehicles, it
will be necessary to understand precisely the interactions between
them.  Indeed, we can assume {\it a priori} that the system will
self-organise itself to some attractor, which may or may not be
a desirable final state, from the point of view of the end-user.
The final emergent behaviour is likely to be complex in nature,
bearing little or no similarity to the original {\it rules} under
which each vehicle proceeds.  Indeed, the scope of this research
is to establish a phase diagram for the system; how the system
globally behaves for given rule-sets applied to each vehicle.
It is our intention to study the transport, navigation,
routing and communications algorithms and infrastructure.

Heavy traffic in one direction will cause a street to be principally
used by vehicles travelling in that direction.  The real-time
peer-to-peer propogation of information will allow other vehicles
travelling in the opposite direction to adjust their routes accordingly.

\section*{Details of the model}

The model discussed in this paper consists of a single multi-lane
street, which is the first element of a larger simulation which will
eventually encompass a small city.  Lanes have no intrinsic directionality
and vehicles are inserted at either end and any lane at random, subject
to that lane being unoccupied in the vicinity of the entrance.
All vehicles behave identically within their parameter specifications.

A central concept to this model is that of the {\it weight} of a
vehicle, with a default value of unity.  "Heavier" vehicles
may claim right-of-way over less heavy vehicles.
In order to resolve conficts between vehicles of equal weight,
each weight is adjusted by a small random value.  Vehicles travelling
in the same direction may form {\it trains} by summing their weights,
and so obtain more right-of-way from oncoming vehicles in the same
lane.

As oncoming vehicles much somehow pass one another, even on single
lane streets, the ruleset assumes that any vehicle which comes to
a halt also accosts itself to one side, and no longer blocks the
lane.

All vehicles have a "sensor" which collects all necessary information
about other vehicles in the vicinity.  The range of this "sensor" is
a parameter to the program, though we have observed that it has little
effect except when decreased below the vehicles' typical safe stopping
distance.

\begin{figure}
\centering 
\includegraphics[scale=0.9]{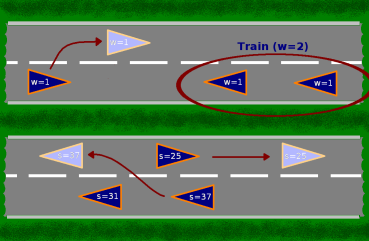}
\caption{\label{ruleset} A simple illlustration of two of the rules which vehicles obey. 
On top, the train arriving from the right, with combined weight $w=2$, obliges the single vehicle
of weight $w=1$ to change lane (or to stop if no other lane is available).  Conflicts between
equal weight vehicles or trains are resolved by random numbers.  The bottom example
shows how a vehicle of speed $s=37$ waits until the lane is clear before moving to pass
a slower vehicle in front (if no lane were available, it would slow down and form a train).}
\end{figure}

Figure~\ref{ruleset} illustrates two of the rules which vehicles obey.
Given that each vehicle "knows" exactly how all other vehicles behave,
collision avoidance is simply a matter of having compatible rules for
all vehicles involved in any interaction.  Once this basic requirement is
satisfied, one may move to study the properties of the model.

\section*{Results}

The simulations we have run to date on the model have revealed that there
is much work to be done.  Even the few parameters specifying the model
permit a very large variation in subsequent behaviour.  We outline some
aspects of this behaviour here.  For all results reported below, we have
fixed the number of lanes at 4, the length of the road at 2500 m, and,
except where specified, the maximum velocity of each vehicle is
a random number uniformly distributed from 10 to 30 m/s.

\begin{figure}
\centering 
\includegraphics[scale=0.4]{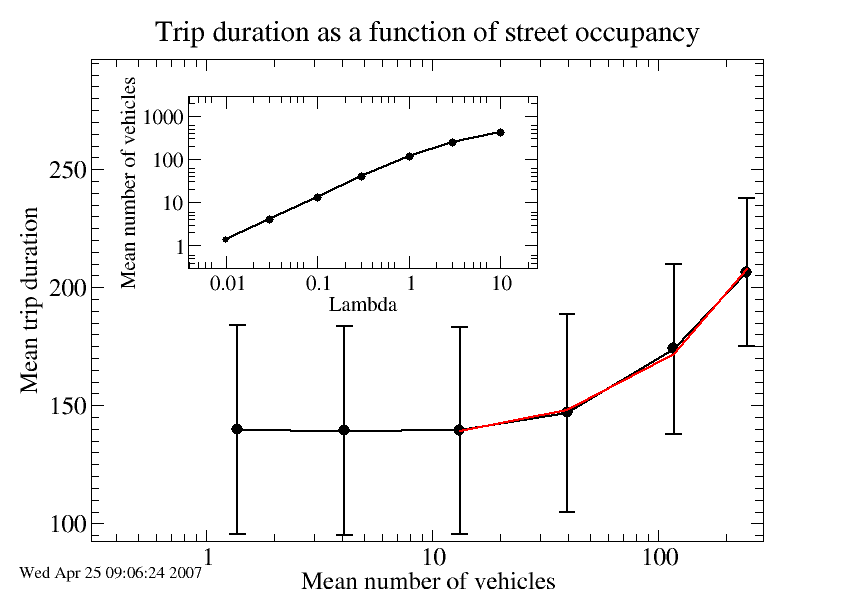}
\caption{\label{transit-vs-ndrones} The mean transit time as a function of the total
street occupancy.  For very low occupancy, the mean transit time is constant, as vehicles
do not interact.  Above a certain threshold, transit time increases roughly linearly.
The inset show the occupancy of the street as a function of $\lambda$, the probability
per unit time that the simulation seeks to insert a new vehicle.  For high $\lambda$
the entrance points might be occupied and the linear relationship at low $\lambda$
is no longer valid.}
\end{figure}

Figure~\ref{transit-vs-ndrones} shows the variation of the average
journey time as a function of the mean number of vehicles $N_v$ on the street.  This
parameter is altered by adjusting $\lambda$ the probability per unit time with which the
system seeks to insert a new vehicle in each timestep.  From the
inset it is clear that
when $\lambda\simeq 0$, $N_v\propto \lambda$.  As $lambda$ increases, the entrance
points to the street saturate and additional vehicles find it difficult to enter.
The main figure instead demonstrates that the mean journey time remains constant
for $N_v\to 0$, but, as the street begins to fill and vehicles find themselves
impeded by those in front, the journey time increases
approximately linearly with $N_v$.

\begin{figure}
\centering 
\includegraphics[scale=0.4]{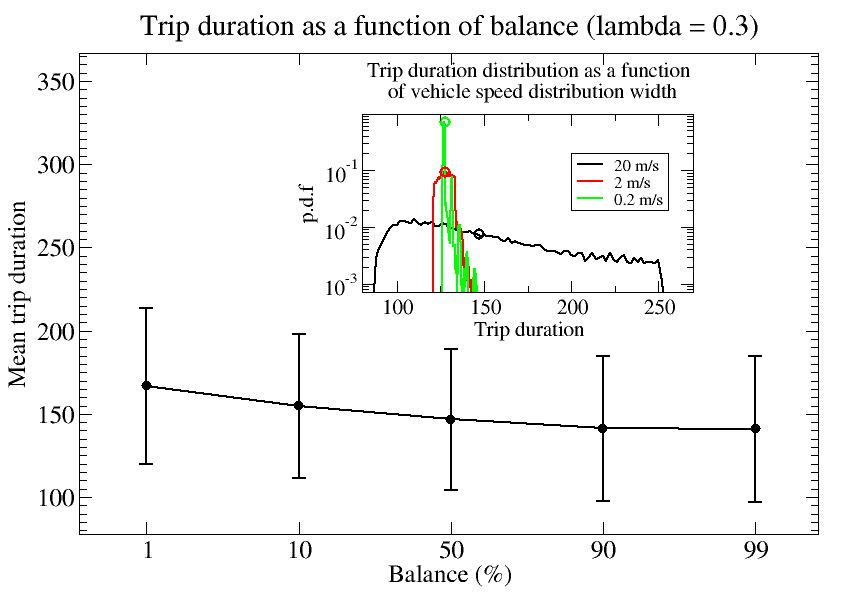}
\caption{\label{transit-vs-balance} The mean transit time as a function of the percentage
of traffic coming from a given direction.  At low values vehicles travel against
the majority and transit time is increased, while, as one might expect, travelling with
the flow is advantageous.  The inset shows the transit time distribution  for three
values of the input speed distribution width.  Clearly, identical vehicles have a positive
effect of the system's overall efficiency.}
\end{figure}

We have adapted the simulation to allow for unbalanced traffic, when
vehicles arrive predominantly from one end or the other of the street.  
Figure~\ref{transit-vs-balance} shows the mean journey time
for cars entering from the left as a function of the percentage of total
cars they represent (i.e. 1\% implies most cars came from the right, 99\%
implies from the left).  This was performed at $\lambda=0.3$ (at the point
where vehicle interaction begins to impact on transit time)
and it is clear that passively imposing a directionality
on the street in this manner improves the journey time for those travelling
with the main flow but penalises those travelling against.
The inset to this figure demonstrates how the transit time varies with
the homogeneity of the vehicles (at $\lambda=0.3$).
When the width of the uniform distribution
from which vehicle speeds are drawn is reduced from 20 to 0.2 m/s, while
maintaining a mean of 20 m/s, the transit time distribution narrows almost
to a delta, and reduces it's mean by some 15\%.

\begin{figure}
\centering 
\includegraphics[scale=0.4]{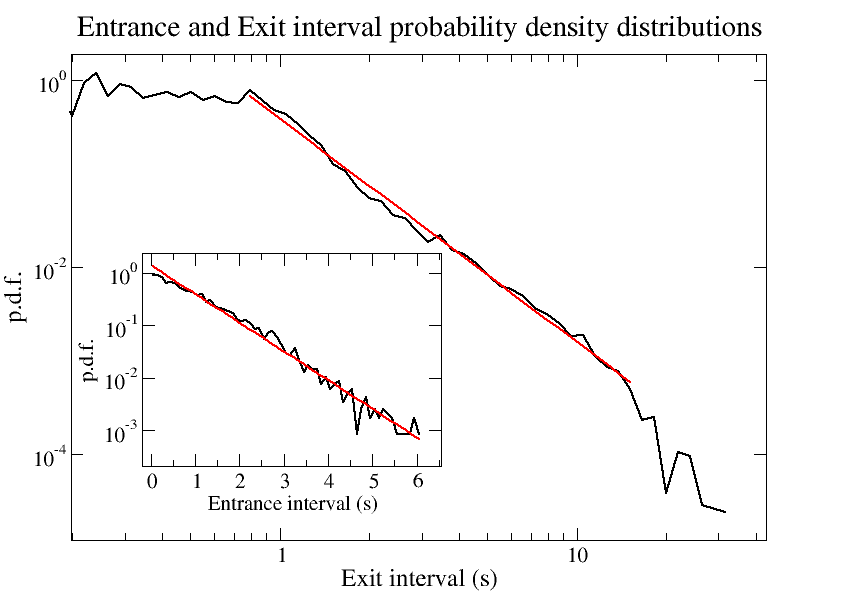}
\caption{\label{interval-pdf-ran} The main figure shows the probability density of
the interval between successive vehicles exiting the street.  There is a scale-free
power-law region between 1 to 10 seconds.  The inset indicates the exponentially
decaying probability of having a given interval between the entrance of two successive
vehicles.  }
\end{figure}

\begin{figure}
\centering 
\includegraphics[scale=0.4]{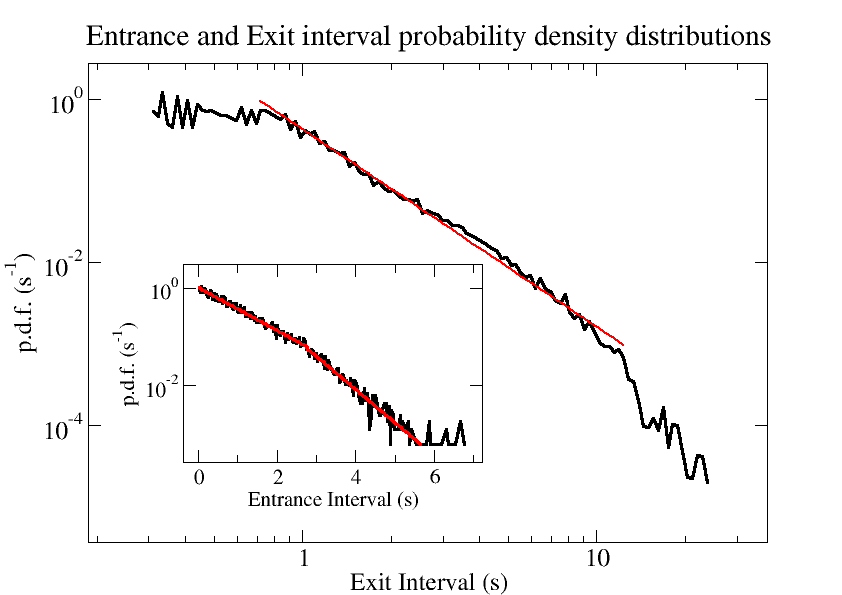}
\caption{\label{interval-pdf-left} The main figure shows the probability density of
the interval between successive vehicles exiting the street when the "keep left"
rule is applied.  The scale-free power-law region between 1 to 10 seconds is still
evident.  The inset indicates the exponentially
decaying probability of having a given interval between the entrance of two successive
vehicles.  }
\end{figure}

We now consider the processes of vehicles entering and exiting the street.  As
the vehicles enter with a given probability in each timestep, we expect this 
to be a Poisson process, with an exponentially improbable interval between
the insertion of two subsequent vehicles.  The inset for figure~\ref{interval-pdf-ran}
demonstrates indeed that this is the case, while the main figure shows the
same distribution for the exit process.  Curiously the exit interval follows
a power law from 1 to 10 seconds after which it decays rapidly.  This
suggests that vehicles are clustering in the street, and indeed, the screenshot
shown in appendix~\ref{snapshot} clearly indicates that this is so.
That vehicles should exit with a power-law distribution of exit intervals is
not desirable from a transport point of view, so it will be interesting to
see how this feature behaves when many streets are modelled together.
Figure~\ref{interval-pdf-left} shows the same distributions
for the "keep left" model; diuscussed below.

As the simulations presented here all specify a street with 4 lanes, we
thought it interesting to see how the lanes are utilised by the vehicles.
Our objective in this preliminary paper is to verify if vehicles can
be made to form "trains" with a minimum of rules, and so efficiently
utilise the infrastructure available.  To visualise how the lanes are
being utilised, we assign a "direction" to each lane which is equal
to the number of vehicles moving in the positive direction minus the number
moving in the negative direction, in that lane.

\begin{figure}
\centering 
\includegraphics[scale=0.4]{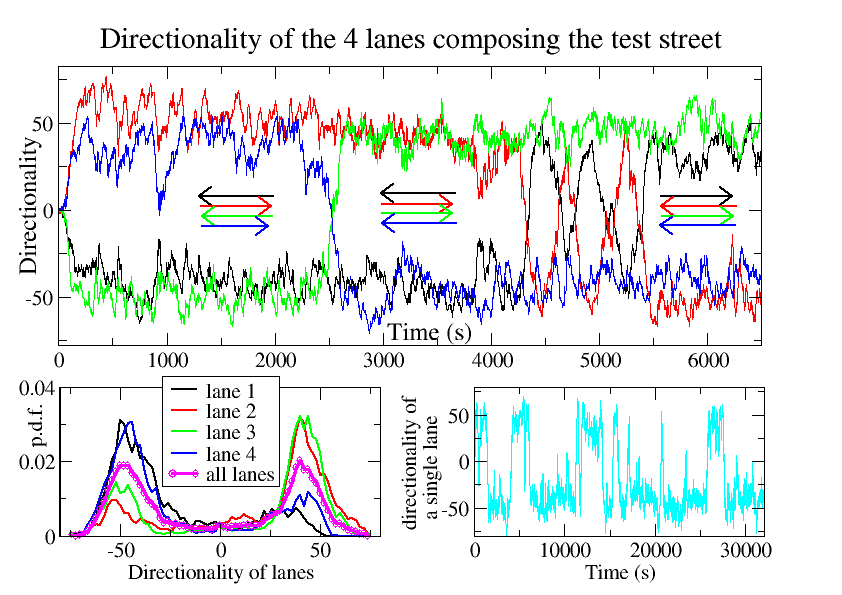}
\caption{\label{lanes-ran} The directionality acquired by each lane as a function of
time.  The bottom left graph shows the probability density of a given directionality,
for each lane individually and for all lanes together.
The bottom right graph shows the directionality of a single lane.  }
\end{figure}

Figure~\ref{lanes-ran} demonstrates that lanes do passively acquire
a directionality.  The main graph shows the direction of each lane as
a function of time - it may be seen that lanes 2 and 4 immediately
established a stable positive direction and 1 and 3 negative.  After some time,
lanes 3 and 4 exchange directions,
and so on.  The bottom left graph shows the probability distribution of each
lane having a given direction.  Though lanes may swap from positive to negative
and vice-versa, evidently they always remain "polarised".  The bottom right
graph shows the directionality of a single lane as a function of time, discussed
below.

Considering figure~\ref{lanes-ran} one may observe that
adjacent lanes frequently acquire opposite directions.  Upon
reflection, we considered that this reduces efficiency, as faster
vehicles will find it difficult to change to a different lane in
order to pass slower vehicles.  We have therefore implemented a
"keep left" rule in which, whenever a vehicle is required to
change lane, it automatically seeks first to move to its left,
choosing the right only if the left is not available.  Furthermore,
a vehicle proceeding normally will always seek to move to its left
if possible.  Naturally, our expectation was that lanes 1 and 2
would acquire a stable positive direction, with lanes 3 and 4 negative.

\begin{figure}
\centering 
\includegraphics[scale=0.4]{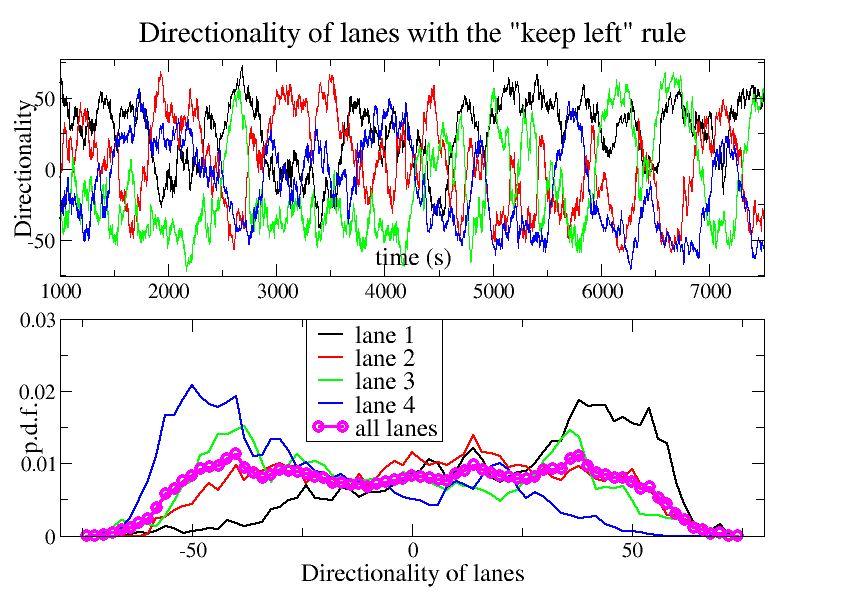}
\caption{\label{lanes-left} The directionality acquired by each lane as a function of
time when the "keep left" rule is applied.
The bottom graph  shows the probability density of a given directionality,
for each lane individually and for all lanes together.}
\end{figure}

Figure~\ref{lanes-left} indicates however, that this does
not occur.  In fact, the "keep left" rule clearly has an undesirable
effect: lanes lose much of their directionality, and change from
positive to negative considerably more frequently.
The "keep left" rule did not cause vehicles to keep very much
to the left!  This is a very simple example of the complexity
inherent to the problem: though individual, isolated
vehicles will stay on the left, many interacting vehicles do not.

\begin{figure}
\centering 
\includegraphics[scale=0.4]{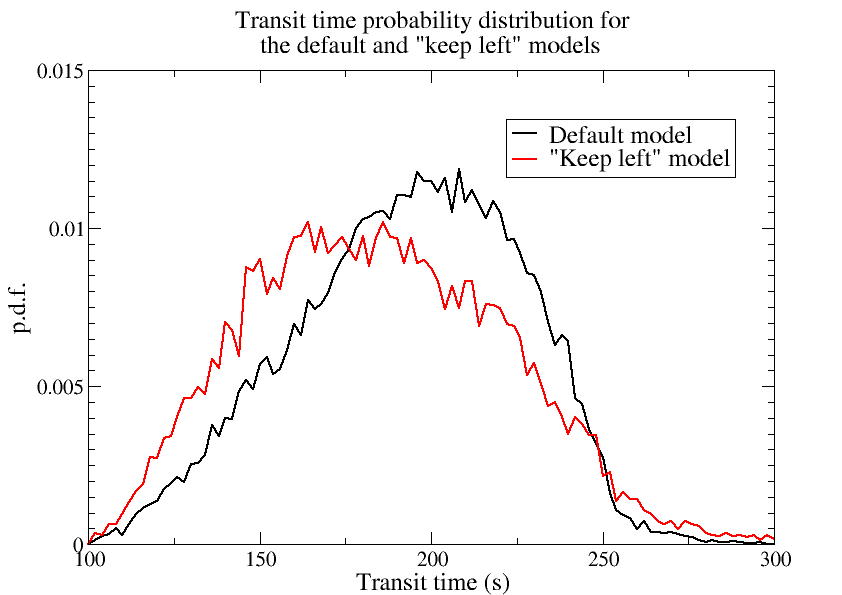}
\caption{\label{ran-left-transit-compare} A comparison between the transit time
probability density distribution for the default and "keep left" models.  "Keep left"
slightly reduces the mean value, but increases the width, and greater statistics are
necessary to establish if the tails of the distribution are unfavorable in either one
case or the other. }
\end{figure}

Strangely, however, the keep left rule does in fact improve
the overall efficiency.  Mean transit time for all vehicles
improves from 195 to 185 seconds, though the width of the
distribution increases by appx.\ 6 seconds and the tail
seems longer.  Figure~\ref{ran-left-transit-compare}
shows a comparison between the duration distrubtions for two
realisations, one with the "keep left" rule, one without.  This
distribution do not seem to follow any particular form, though
we have observed that it is highly dependent on the ruleset.
Figure~\ref{interval-pdf-left} shows the
entrance and exit interval distributions for the "keep left" model;
they do not significantly change with respect to the default,
though it seems that the Poisson entrance process is somewhat
modified - it may be that vehicles clustering on the left near
the entrance is the cause.  It yet remains to be verified if the
"keep left" rule actually confers a net advantage or not.


This article represents the first results in what is planned
to be a much more comprehensive simulation of an autonomous
transport system.  As such, it is very much exploratory and
the meaning of these first observations have yet to be
defined and categorised in accordance with existing research.

Nonetheless, we note some interesting points.  That
the exit process follows a power-law distribution of intervals
suggests that vehicles are somehow joining into
scale-free clusters.  Clearly these "clusters" are fundamentally
different to typical "traffic jams" of the day: traffic jams are
clusters of stationary vehicles, while our trains are clusters of
moving vehicles.  Nonetheless, the clusters generated here are naturally
limited by the length of the street, and so it will be necessary
to ensure they do not diverge as the model grows in complexity
and size.

Some of the results we have also bear interesting similarities
to other physical systems.  For example, in figure~\ref{lanes-ran},
bottom-right graph, the directionality of a single lane
bears some similarity to the magnetisation of the Ising model as
it fluctuates in time.  Further analysis is necessary to verify
if the similarity is mathematically consistent or merely superficial.
Certainly the model presented here, in
the limit of many lanes, could effectively act as an Ising model
with the directionality as a spin.  This opens
up the fascinating question as to whether an energetic argument could
be brought to bear upon the simluations - for example a vehicle
will change lane, against the flow of traffic, only if it can "borrow"
sufficient energy from a reservoir.  We note, however, that such
an argument would not of little relevance for real traffic, as lanes,
justifiably, have an intrinsic, constant direction.

Furthermore, traffic has been described as a "Self-Organised Critical"
system~\cite{bak-how-nature-works} and it would certainly be
revealing if this model were found to be critical as the
Ising model is known to be.

We have simulated the flow of automated traffic on a 4-lane street,
in which lanes have no directinality.  Given the ruleset imposed,
we find that vehicles can be made to form clusters of
"trains" which efficiently utilise the street.  The model
seems to behave as a "complex system" in that we have observed
some scale-free behaviour, and, though the system presented here is
quite small, its overall behaviour is
not trivially linked to the basic ruleset governing the vehicles.

Our objective in this article was to satisfactorily simulate a
single street as a basic building block for a larger model, eventually
encompassing perhaps a small city.  We consider that, even at this
basic level, there is much analysis to be done, and finding the optimum
ruleset for the autonomous vehicles will be far from trivial.


\section*{Acknowledgments}
This work has been supported by the EU 6th Framework Programme, under contract number 044386,
titled "TRIGS - Triggering of Instabilities in Materials and Geosystems" and by the Italian
"Fondo per gli Investimenti della Ricerca di Base" under contract nr. RBAU01883Z, titled
"Propriet\'a Reologiche di Materiali Granulari".

\newpage

\begin{figure}
\centering 
\includegraphics[scale=0.4]{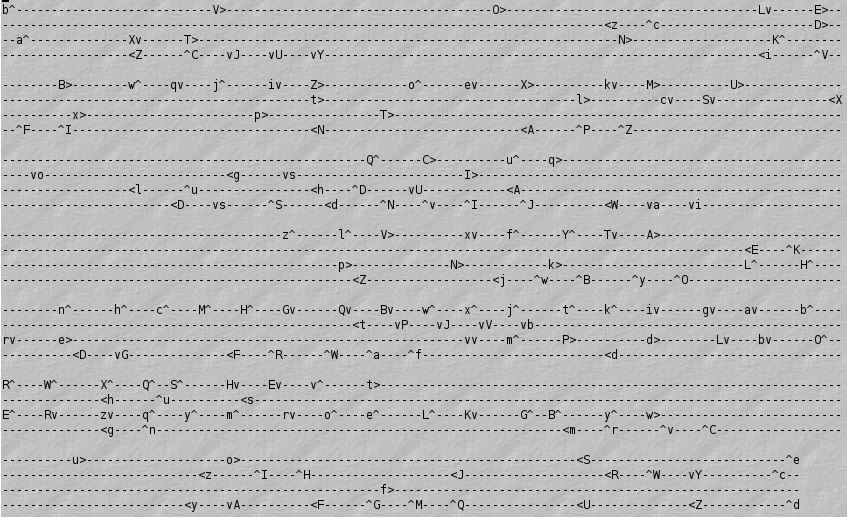}
\caption{\label{snapshot} A screenshot of the street in an "ncurses" window.  Each vehicle
is denoted by a single letter, and its state (moving, accelerating, stopped) by
an additional character.  The simulation also allows for collisions which fortunately
seem not to occur after a long process of debugging!  }
\end{figure}

\end{document}